\documentclass[final,authoryear,3p,arial,onecolumn]{elsarticle}

\usepackage{graphicx,amssymb,amsmath,amsfonts,bm}
 \usepackage{epstopdf}
\usepackage{color}
\usepackage{graphicx}
 \usepackage{mathptmx}      
\usepackage{latexsym}
\usepackage{amssymb}

 \definecolor{olive}{rgb}{0.333333,0.419608,0.184314}
\definecolor{brown}{rgb}{0.737255,0.560784,0.560784}

\usepackage{graphics}
\newcommand{\trademark}[1]{#1\textsuperscript{\textregistered}}
\usepackage{subfigure}
\usepackage{wasysym}
\usepackage{latexsym}
\usepackage{amssymb}

\ifpdf
  \usepackage[pdftex]{hyperref}
\else
  \usepackage[hypertex]{hyperref}
\fi

\hypersetup{backref,
  a4paper,
  colorlinks=true,
  urlcolor=red,
  raiselinks=true,
  linkcolor=red,
  runcolor =red,
  citecolor=red,
  pdfpagemode=FullScrren,
  pdftitle={Thermo- Carbopl},
  pdfauthor={Teodor Burghelea},
  pdfcreator={$ $ Teodor I. Burghelea$ $},
  pdfsubject={Thermorheology Carbopol},
  bookmarks=true
  pdfkeywords={}
}

\journal{Journal of Non Newtonian Fluid Mechanics}

\begin{document}

\begin{frontmatter}

\title{Unsteady laminar pipe flow of a \trademark{Carbopol} gel. Part I: experiment.}

\author{Antoine Poumaere}
 \address{LUNAM Universit\'{e}, Universit\'{e} de Nantes, CNRS, Laboratoire de Thermocin\'{e}tique, UMR 6607, La Chantrerie,  Rue Christian Pauc, B.P. 50609, F-44306 Nantes Cedex 3, France}
 
\author{Miguel Moyers-Gonz\'alez}
\ead{Miguel.MoyersGonzalez@canterbury.ac.nz}

\address{Department of Mathematics and Statistics, University of Canterbury, Private Bag 4800, Christchurch, 8041, NZ }

\author{Cathy Castelain}
 \address{Cathy.Castelain@univ-nantes.fr}

\author{Teodor Burghelea}
\ead{Teodor.Burghelea@univ-nantes.fr}

\address{LUNAM Universit\'{e}, Universit\'{e} de Nantes, CNRS, Laboratoire de Thermocin\'{e}tique, UMR 6607, La Chantrerie,  Rue Christian Pauc, B.P. 50609, F-44306 Nantes Cedex 3, France
}%

\begin{abstract}

A experimental study of low Reynolds numbers unsteady pipe flows of a yield stress shear thinning fluid (\trademark{Carbopol} - 980) is presented.  The investigation of the solid-fluid transition in a rheometric flow in the presence and in the absence of the wall slip reveals a coupling between the irreversible deformation states and the wall slip phenomenon. Particularly, the presence of wall slip nearly suppresses the scaling of the deformation power deficit associated to the rheological hysteresis with the rate at which the material is forced. The irreversible solid-fluid transition and the wall slip behaviour emerge in the same range of the applied stresses and thus, the two phenomena appear to be coupled to each other. In-situ measurements of the flow fields performed during an increasing/decreasing stepped pressure ramp reveal three distinct flow regimes: solid (pluglike), solid-fluid and fluid. The time resolved measurements of low Reynolds number unsteady flows driven by an increasing/decreasing controlled pressure ramps reveal a irreversible yielding scenario similar to that observed in the rheometric flows. The deformation power deficit associated with the hysteresis observed during the increasing/decreasing branches of the pressure ramps reveals a dependence on the rate at which the unsteady flow is driven consistent with that observed during the rheological measurements in the presence of slip.  
The dependence of the slip velocity on the wall shear stresses reveals a Navier-type slip behaviour only within the fluid flow regime, which indicates that the wall slip phenomenon is directly coupled to the solid-fluid transition. A universal scaling of the slip velocity with the wall velocity gradients is found and the slip length is independent on the characteristic time of forcing $t_0$. 
The paper closes with a discussion of the main findings, their possible impact on our current understanding of the yielding and slip behaviour of \trademark{Carbopol} gels. Several steps worth being pursued by future experimental/theoretical studies are proposed.

\end{abstract}




\end{frontmatter}

\clearpage
\tableofcontents
\listoffigures
\clearpage

\section{\label{sec:intro}Introduction}

Yield stress fluids represent a broad class of materials made of high molecular weight molecular constituents which display a solid-like behaviour as long as the stress applied onto them does not exceed a critical value called the yield stress, $\tau_y$, and a fluid one beyond this threshold.  

The constantly increasing level of interest of both theoreticians and experimentalists in yield  stress fluids has, in our opinion, a two-fold motivation. From a practical perspective, such materials have found an increasing number of practical applications for several major industries (which include foods, cosmetics, oil field, etc.) and they are encountered in the daily life in various forms such as food pastes, hair gels and emulsions, cement, mud and more. 
\trademark{Carbopol} gels have been widely used in many rheological and hydrodynamic experimental studies and they have been considered for over a decade a \textit{model yield stress fluids}. The wide use of \trademark{Carbopol} gels is equally due to their optical transparency which allows an in-situ assessment of the flow fields, their stability (they can be stored for extended periods of time without noticeable changes in their rheological behaviour), \cite{piaucarbopol}. \trademark{Carbopol} resins are synthetic polymers of acrylic acid initially introduced over six decades ago (B.F. Goodrich Co.). They are cross-linked with various chemical compounds such as divinyl -glycol, allyl-sucrose, polyalkenyl polyether etc. In an anhydrous form, the average size of the polymer molecules is of the order of hundreds of nanometers. In the absence of cross-links, the polymer particle can be viewed as a collection of linear chains intertwined (in a coiled state) but not chemically bonded. The \trademark{Carbopol} micro-gel particles are soluble in polar solvents and, upon dissolution, each individual polymer molecule hydrates, partially uncoils, and swells several orders of magnitude. Addition of a neutralising agent leads to the creation of negative charges along the polymer backbone due to ionisation of the carboxylic acid groups. Consequently, swollen polymer molecules cross-link, forming a system of micro-gel particles. The micro-gel system can sustain finite deformations (behaving like an elastic solid) prior to damage. When local applied stresses exceed a threshold value the gel system breaks apart and the material starts to flow. This is the commonly accepted microscopic scale origin of the macroscopic yielding of the material.
 
From a fundamental perspective and even in the case of minimal thixotropic effects, yield stress materials continue triggering intensive debates and posing difficult challenges to both theoreticians and experimentalists. From a historical perspective, the best known discussion is related to the very existence of the yield stress \cite{barnes1, barnes2}.
It was thus suggested that the yield stress is a \textit{engineering} reality rather than a physical one or, in other words, the apparent yield stress behaviour observed during rheological tests is related to a very large Newtonian viscosity rather than to the onset of flow (yielding).  

In context of the recent developments of the rheometric equipment (now able to resolve torques as small as $0.1 nNm$  and rates of shear as small as $10^{-7}~s^{-1}$), a further pursue of this already classical debate on the existence of a yield stress is, in our opinion, trite. Indeed, recent rheological tests performed on various \trademark{Carbopol} gels proved unequivocally the existence of a true yielding behaviour, \cite{solidfluid, moller1, dennyieldstress, dennrheoacta}. 

The existence of a Hookean solid deformation regime preceding the solid-fluid transition has been experimentally demonstrated in Ref. \cite{solidfluid} in the form of a plateau of the rate of deformation measured during increasing/decreasing linear stress ramps. The following characteristics of the transition from a solid like to a fluid like behaviour in a rheometric flow (a plate-plate geometry) in the absence of wall slip have been found experimentally:

\begin{enumerate}
\item{The solid-fluid transition is not direct but mediated by an intermediate deformation regime characterised by a coexistence of the solid and fluid behaviour. This experimental facts render the classical Herschel-Bulkley formalism inapplicable over the entire range of material deformations.}

\item{Except for the fluid deformation regime observed for applied stresses significantly larger than the yield stress $\tau_y$, the solid-fluid transition is not reversible upon increasing/decreasing the applied stress and a non-negligible hysteresis effect is observed.}

\item{In the absence of wall slip, the deformation power deficit associated to the experimentally observed hysteresis scales as a power law with the rate at which the material is deformed, i. e. the inverse of the time interval the applied stress is maintained constant during the increasing/decreasing branches of the stress ramp. Therefore, only in the asymptotic limit of very slow deformations the hysteresis tends to disappear and the solid-fluid transition becomes reversible.}
\end{enumerate} 
 
 The experimentally found features of the solid-fluid transition in a \trademark{Carbopol} gel enumerated above cannot be captured within the classical theoretical frameworks (Bingham, Herschel-Bulkley, Casson) which predict a reversible deformation and a well defined onset of flow.

 The rheological characterisation of several structurally and physico chemically different yield stress fluids (a  \trademark{Carbopol} gel, a commercial hair gel, a commercial shaving foam and a cosmetic water-in-oil emulsion) presented by M\o{}ller an his coworkers bring additional strong arguments for the existence of the yield stress and seems to rule out the possibility of a large Newtonian viscous behaviour prior to yielding,  \cite{bonn_europhys}. Whereas a viscosity plateau is indeed observed within a range of low applied stress, its value increases as a power law with the characteristic time $t_0$  during each the shear rate data is averaged for each value of the applied stress, $\eta_0 \propto t_0 ^ {\alpha}$ (with $\alpha \approx 1$). Thus, the authors conclude that the apparent viscosity plateau is only due to the fact that a steady state of deformation is not achieved for low applied stresses and therefore it should be interpreted with caution (see Fig. 2 in Ref. \cite{bonn_europhys}). The experimentally found power law scaling of the plateau viscosity observed prior to yielding implies a divergence of this viscosity which M\o{}ller and his coworkers interpret as an unequivocal proof for the existence of a true yield stress.

The insufficiency of the classical descriptions of the yield stress fluids has been clearly emphasised by Denn and Bonn, \cite{dennrheoacta, dennyieldstress}. To overcome these difficulties, Denn and Bonn propose in Ref. \cite{dennrheoacta} several ingredients which a realistic theoretical model should contain: \textit{thixotropy, avalanche behaviour, loss of fore-aft symmetry in flow, shear banding without a stress gradient, shear localisation}. At a technical level, they equally suggest employing a kinetic equation for a structural parameter which dynamically characterises the amount of un-yielded material as a function of the externally applied stress, see Eq. 7 in Ref. \cite{dennrheoacta}.  To these ingredients we would also add the practical need of accounting for the wall slip behaviour. 

Part of these tasks (which in our opinion are fully justified by the compelling experimental evidence) have already been addressed (at least partially). A simplistic kinetic model that reveals a thixotropic behaviour (time dependent yielding, irreversibility of deformation states), can phenomenologically explain the fore-aft symmetry breaking during flow past moving rigid body observed experimentally in Ref. \cite{sedimentation} and is consistent with a shear banding (phrased as coexistence between solid and fluid behaviour) has been proposed in Ref. \cite{solidfluid}. Although this model proved its ability to describe the solid-fluid transition in a rheometric flow (e.g. it can accurately fit stress strain data acquired during controlled stress ramps), additional theoretical developments are needed in order to describe more complicated flow situations such as sedimentation problems, start-up pipe flows etc. We also note that this oversimplified model was not formulated from first principles and it is still lacking a thermodynamic validation. 

Whereas the recent experimental evidence for the existence of yield stress seems compelling to us, we do believe, however, that dynamics of the yielding transition remains an interesting topic for both theoreticians and experimentalists, particularly in realistic flow settings where a steady state of deformation can not be achieved and in the presence of slippery boundaries. In practical flow situations involving pasty materials the yielding transition is often associated with the wall slip phenomenon which complicates even further the theoretical description. Contrarily to the case of polymer melts where the slip is observed at large rates of strain \cite{dennslipreview, dennbook}, the wall slip  in pasty materials emerges within a range of small strains. This implies that in the case of pasty materials the wall slip and the yielding transition are most probably coupled to each other which brings an additional experimental and theoretical difficulty in exploring the solid-fluid transition and modelling realistic flows of such materials in industrially relevant settings.

A detailed experimental investigation of coupling between the wall slip phenomenon and yielding observed in a coaxial cylinder geometry for various pasty materials has been reported by Bertola and his coworkers, \cite{bertolaslip}. The novelty of their study comes from combining classical rheometry techniques (increasing/decreasing controlled stress ramps) which provides the characterisation of wall slip and yielding at an integral scale with the magnetic resonance imaging (MRI) technique which provides local measurements of the velocity profiles. The successful combination of these complementary techniques  allowed Bertola and his coworkers to provide a detailed description of the coupling between slip and yielding and to identify a critical rate of shear below which no steady state flow fields can be observed by MRI, \cite{bertolaslip}.

A detailed theoretical study of the combined effects of compressibility and slip of Poiseuille flows of a Herschel-Bulkley fluid has been recently presented by Damianou and coworkers, \cite{Damianou2012}. A central conclusion of this work is that, in the incompressible flow case, the flow field may become plug-like corresponding to a critical value of the slip parameter which is found to be proportional to the yield stress of the material. 

Whereas there exists a significant number of studies of the yielding of \trademark{Carbopol} gels in rheometric flows, the body of literature concerning experimental studies of the solid-fluid transition of \trademark{Carbopol} gels in industrially relevant flows is, to our best knowledge, more limited. 
A experimental study of the yielding and flow properties of a $0.2 \%$ \trademark{Carbopol} in a capillary tube has been performed by Gonz\'alez and his coworkers, \cite{pipesteady}. By combining classical rotational rheometry techniques with Digital Particle Image Velocimetry (\textbf{DPIV}) measurements of steady flows driven at various pressure drops the authors of this study conclude that a yield stress behaviour does exist.  
As the driving pressure drop is gradually increased, they observe a smooth solid-fluid transition. To overcome the difficulty of accurately measuring the yield stress triggered by the smoothness of the solid-fluid transition, the authors of the study propose an alternative method of measuring this parameter based on the analysis of the measured velocity profiles, \cite{pipesteady}.
As the flows studied in Ref. \cite{pipesteady} were stationary, these experimental results can neither confirm nor contradict the main features of the solid-fluid transition observed during unsteady rheometric flows in Refs. \cite{solidfluid, thermo} which motivated us to consider the unsteady pipe flow case in the present study.  

The body of existing literature on unsteady pipe flows of \trademark{Carbopol} gels is even more limited. 
 Park and his coworkers have studied both experimentally and theoretically oscillatory flows of a $0.075 \%$ \trademark{Carbopol} solution in a acrylic made U-shaped tube with a circular cross section , \cite{park}. Consistently with the findings reported in Ref. \cite{solidfluid}, the authors of this study have provided a clear experimental evidence on the significant role of elasticity on the measured unsteady velocity profiles particularly near the centreline of the flow channel. Due to the elastic effects, strongly nonlinear time histories of the strain are measured. To model their experimental findings, Park and his coworkers propose an elasto-viscoplastic model which fits their data better than the regularised Bingham model. Some discrepancies between theory and experiment still exist and may be due to the fact that their model does not account for the velocity slip near the wall.  

The central aim of the present study is to present a detailed experimental characterisation of unsteady flows of a \trademark{Carbopol} gel in the presence of wall slip in both a rheometric flow and a pipe flow. The general question addressed through our study concerns with the dynamics of the solid-fluid transition observed during the unsteady low Reynolds number pipe flow of a \trademark{Carbopol} gel and its connection to both the wall slip phenomenon and the solid-fluid transition previously observed in rheometric flows (with and without slip at the boundaries). 

The paper is organised as follows. 
The experimental setup, the measuring technique and the physical and rheological properties of the solutions are detailed in Sec. \ref{sec:experimental}.
The experimental results are presented in Sec. \ref{sec:results}. A systematic description of the solid-fluid transition observed in a rheometric flow with a particular emphasis on the role of wall slip and its coupling to the irreversibility of the deformation states is presented in Sec. \ref{sec:sfrheo}.
A detailed characterisation of inertia-free unsteady channel flows of a \trademark{Carbopol} solution is presented in Sec. \ref{subsec:pipeflow}.
The paper closes with a discussion of our main findings, of their impact on the current understanding of flows of yield stress fluids, Sec. \ref{sec:conclusions}. Several directions worth being pursued by future studies will also be discussed.

\section{\label{sec:experimental}Experimental setup and methods}

\subsection{\label{subsec:setup}Experimental apparatus}

\begin{figure*}
\includegraphics[width=0.75\textwidth]{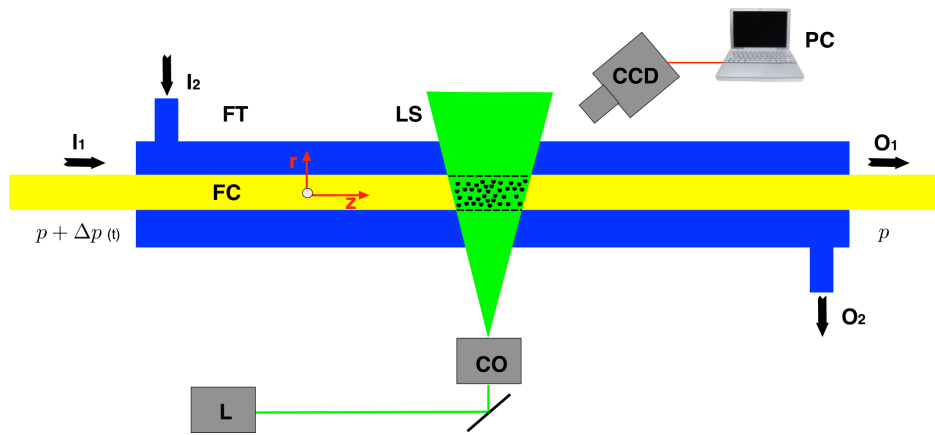}\caption{\label{fig:setup} Schematic view of the experimental apparatus: \textbf{FT} - fish tank, $\bf{I_2}$ - water inlet, $\bf{O_2}$ - water outlet, \textbf{FC} - flow channel, $\bf{I_1}$ - working fluid inlet, $\bf{O_1}$ - working fluid outlet, \textbf{L} - solid state laser, \textbf{CO} - cylindrical optics block, \textbf{LS} - laser sheet, \textbf{CCD}  - digital camera, \textbf{PC} - personal computer.}
\end{figure*}

The experimental setup is schematically illustrated in Fig. \ref{fig:setup}. It consists of a flow channel \textbf{FC} of length $L=115~cm$ with a circular cross-section of inner radius $R=1.9~mm$.  

The flow channel is housed by a acrylic made fish tank \textbf{FT} filled with water circulating at a constant temperature, $T=23 ^o C$.
Besides its role in maintaining a constant operating temperature, the water circulation around the flow channel also provides a good matching of the refractive index which minimises the optical distortions induced by the curved surface of the flow channel. 
 
 A green laser beam ($\lambda = 514~ nm$) with a power of $500~ mW$ emitted by the solid state laser \textbf{L} (from Changchun Industries, Model LD-WL206) is deflected by a mirror through a cylindrical optics block \textbf{CO} which reshapes it into a horizontal laser sheet, Fig. \ref{fig:setup}.
The cylindrical optics block is composed of a glass rod with a short focal distance $f_1 \approx 2  ~ mm$ and a cylindrical lens with a larger focal distance, $f_2 \approx 5  ~ cm$. The two optical elements are mounted orthogonally to each other and in a telescopic arrangement such that the primary horizontal laser sheet generated by the glass rod is focused on the vertical direction in the middle of the flow channel by the second lens minimising thus its thickness in the measurement region. The thickness of the generated laser sheet is roughly $80 \mu m$ in the beam waist region which is positioned at the centreline of the flow channel. The working fluid was seeded with an amount of $200$ parts per million ($ppm$) of polyamide particles with a diameter of $60~\mu m$ (from Dantec Dynamics). A time series of flow images is acquired with a 12 bit quantisation digital camera  \textbf{CCD} (Thorlabs, model DCU224C) through a $12X$ zoom lens (from La Vision). The resolution of the acquired images is $1280X1024 $ pixels which insures a spatial resolution of $2.96 \mu m$ per pixel. 

\subsection{\label{subsec:preparation}Preparation of the \trademark{Carbopol} solutions}

A $0.08 \% $ (by weight) solution of \trademark{Carbopol} $980$ has been used as a working fluid.
The  procedure for the preparation of the solution is described as follows. First, the right amount of anhydrous \trademark{Carbopol} $980$ has been gently dissolved in water while continuing stirring the mixture with a commercial magnetic stirring device.
The stirring process was carried on until the entire amount of
polymer was homogeneously dissolved. A particular
attention has been paid to the homogeneity of the final mixture, which has been
assessed visually by monitoring the refractive index contrast of the mixed solution. Next, the $pH$ of
the mixture (initially around $3.2$, due to the dissociation of the polyacrylic acid in water) has been brought to a neutral value by addition of about $140$ parts per million ($ppm$) of sodium hydroxide ($NaOH$). The final value of the $pH$ has been carefully monitored using a digital $pH$- meter (from Grosseron). Upon the ionisation and the neutralisation of the carboxylic groups of the polyacrylic acid, the \trademark{Carbopol} particles swell dramatically forming a percolated micro-gel, \cite{piaucarbopol}.  As a particular attention has been paid to control the neutral $pH$, the concentration of the sodium ions was presumably equal to that of the carboxylate groups. 
 To homogenize the gelled mixture, the sample has been stirred with a
propeller mixer (at $400~rpm$) for about two hours and then
allowed to rest for several more hours. Prior to each rheological
test, the air bubbles entrapped into the gel during the stirring
process have been removed by placing the samples in a low pressure
chamber for $30$ minutes. Finally, the fluid batch
has been stored at room temperature in a sealed container in order
to minimize the evaporation of the solvent. 

\subsection{\label{subsec:rheological}Rheological characterisation of the \trademark{Carbopol} solutions}

The rheological properties of the solutions were investigated
using a  Mars $III$ (from Thermofischer) equipped with a Peltier system able to control the temperature with an accuracy better than $0.1 ^{\circ} C$ and a nano-torque module which allows one to accurately explore very small rates of deformation, $\dot \gamma \approx 10^{-5}~s^{-1}$. 

 A first major concern for the rheological measurements was the emergence of the wall slip phenomenon at the contact with the measuring geometry, which has been previously recognised as a major source of experimental artefacts when \trademark{Carbopol} gels are used, \cite{piaucarbopol, piau-taosfpvbfaspv2004}. To quantitatively assess the role of the wall slip, two alternative geometries have been used: a smooth parallel plate system and a serrated one. The radius of each geometry was $R=35~mm$ and the gap between the parallel plates was $d=1~mm$. Measurements of flow curves performed with the rough parallel plate system for various gaps indicated that the plate's roughness did suffice to prevent the wall slip (data not shown here).

A second concern was related to the possible artefacts introduced by fluid evaporation during long experimental runs. In order to prevent this, a solvent trap has been placed around the free fluid meniscus. The sealing of the solvent trap on the base plate of the rheometer has been insured by a thin layer of vacuum grease.  After each experimental run it has been carefully checked (by visual inspection) that no significant changes in the shape of the meniscus occurred. Additionally, we have checked at the end of each run that one can reproduce the viscosity measured during the pre-shear step which indicated us that the evaporation effects were either minimal or absent.

\subsection{\label{subsec:flowcontrol}Flow control}

\begin{figure}
\includegraphics[width=1\textwidth]{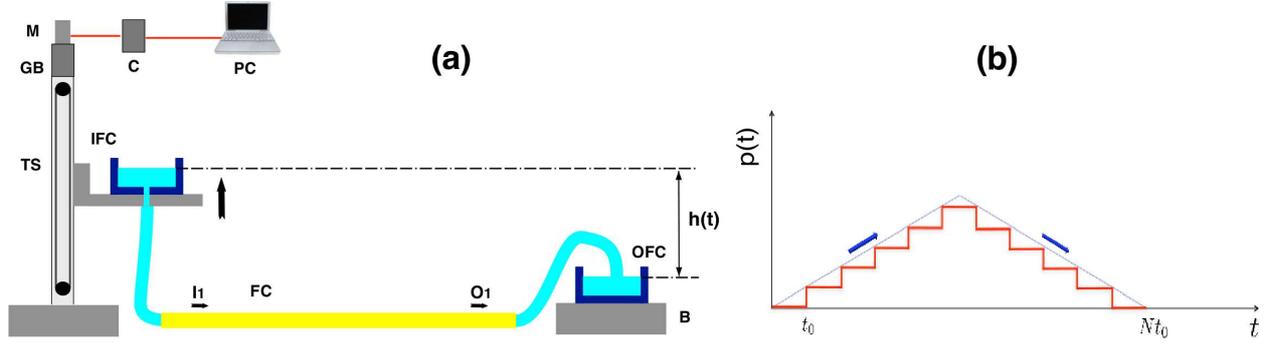}\caption{\label{fig:setup1} \textbf{(a)}  Schematic view of the flow control system: \textbf{TS} - vertical translational stage, \textbf{M} - stepping motor, \textbf{GB} - gear box, \textbf{GB} - micro step controller, \textbf{PC} - personal computer, \textbf{IFC} - inlet fluid container, \textbf{OFC} - outlet  fluid container, \textbf{FC} - flow channel, \textbf{FC} - flow channel, $\bf{I_1}$ - working fluid inlet, $\bf{O_1}$ - working fluid outlet, \textbf{B} - digital balance. \textbf{(b)} Schematical representation of the hydrostatic pressure ramp. $t_0$ is the characteristic forcing time (the averaging time per stress point) and $N$ is the total number of steps.}
\end{figure}

A slow controlled pressure flow has been generated by controlling the hydrostatic pressure difference between the inlet and the outlet of the flow channel as schematically illustrated in Fig. \ref{fig:setup1}. The working fluid is held in a container \textbf{IFC} with a large free surface rigidly mounted on a vertical translational stage \textbf{TS}. The translational stage is driven by a step motor \textbf{M} via a gear box \textbf{GB}.

The position and the speed of the stepping motor are controlled by a micro-step controller (model IT 116 Flash from Isel Gmbh) connected to a personal computer \textbf{PC} via a $RS~232$ serial interface. To control the motion of the stepping motor, a graphical user interface (GUI) has been developed under \trademark{Matlab}. The vertical position of the inlet fluid container \textbf{IFC} is controlled with an accuracy of $37~\mu m$.
After passing through the channel, the fluid is collected in a outlet fluid container \textbf{OFC} placed on a digital balance \textbf{B}. The balance provides real time reading of the mass of the fluid discharged through the channel via a $RS~232$ serial port interfaced under \trademark{Matlab}.

The pressure driving the pipe flow is $p(t)=\rho g h(t) - p_t$ where $\rho = 1050~kg~m^{-3}$ is the density of the \trademark{Carbopol} solution, $g$ the acceleration of gravity and $p_t$ the pressure drop in the connecting tubes. 
Measurements of the flow rate obtained from the mass of the discharged fluid (the data are not shown here) allowed one to estimate that the pressure drop along the connecting tubes $p_t$ roughly accounts for $5 \%$ of the total hydrostatic pressure difference. 

The imposed hydrostatic pressure mimics the flow ramps employed in Ref.  \cite{solidfluid} to investigate the yielding phenomenon and consisted of linear hydrostatic pressure ramps for both increasing and decreasing values of the height difference $h(t)$ (see Fig. \ref{fig:setup1} (b)):
\begin{equation} \label{eq:pressure_ramp}
h(t)=\sum_{i=0}^{N-1}C[i]\left(\bm H(t-it_0)- \bm H(t-(i+1)t_0)\right)
\end{equation}
where
\begin{equation}
C[i]=\frac{2 h_{\max}}{(N-1)t_0}
\begin{cases}
i, & i\leq \frac{N-1}{2} \\
N-(i+1), & \text{o.w.}
\end{cases}
\end{equation}
and $\bm H(t)$ is the Heaviside function. 

The pressure ramp described by Eq. \ref{eq:pressure_ramp} and illustrated in Fig. \ref{fig:setup1} (b) consists of $N$ steps each lasting a time $t_0$. As in the case of the stress ramps studied in \cite{solidfluid} $t_0$ will be referred to as the \textit{characteristic forcing time}.

\subsection{\label{subsec:flowmeas}Flow field measurements}

Time series of the velocity fields were obtained by a iterative multi-grid Digital Particle Image Velocimetry (\textbf{DPIV}) algorithm implemented in the house under \trademark{Matlab} \cite{scarano, piv2}. 
For this purpose, simultaneously with the hydrostatic pressure ramp illustrated in Fig. \ref{fig:setup1}(b) a sequence of flow images has been acquired at a frequency of $10$ frames per second (fps).
Corresponding to low values of the driving pressure, several flow images have been skipped in order to increase the inter-frame and maintain the average displacement of the tracer particles in the optimal range of $5$ to $15$ pixels, \cite{piv2, scarano}. Together with this, the smallest interrogation window has been adapted to the mean flow velocity corresponding to each step of the pressure ramp. 
The spatial resolution of the measured flow fields was $99.5 ~\mu m$. 
To account for the significant mean flow speed variation during the controlled pressure ramps illustrated in Fig. \ref{fig:setup1}(b), an adaptive inter-frame between the subsequent images passed to the \textbf{DPIV} algorithm has been used.  Thus, as the pressure drop has been increased, the inter-frame was gradually decreased from $1.5~s$ (corresponding to the lowest value of the pressure drop) to $0.1~s$ (corresponding to the largest value of the pressure drop investigated). This procedure allowed one to maintain the instrumental errors in the desired confidence interval, regardless the mean flow speed.  

The accuracy of the method has been carefully checked by running test measurements for low Reynolds number Poiseuille flows with Newtonian fluids and comparing them with the analytical solution. Thus, it has been checked that using this adaptive \textbf{DPIV} protocol the instrumental error of the measured velocity fields does not exceed $7 \%$ within the entire range of driving pressures we have explored.

\section{\label{sec:results}Experimental results}

\subsection{\label{sec:sfrheo}The solid-fluid transition in a rheometric flow: the role of wall slip}

Prior to characterising the unsteady pipe flow of a \trademark{Carbopol} gel we focus on the solid-fluid transition which we have investigated in detail in \cite{solidfluid} for this material. It is known that \trademark{Carbopol} gels exhibit slip while flowing past smooth surfaces, \cite{piaucarbopol}. Thus, as the flow channel used in this study is glass-made it is highly expected to observe a non-negligible velocity slip at the wall. This motivated us to perform a comparative rheological investigation of the \trademark{Carbopol} gel in two separate geometries with and without wall slip. 
The rheological response of the \trademark{Carbopol} sample is tested during a stepped stress ramp performed at a constant temperature. This stress ramps closely mimics the pressure ramp that drives the channel flow illustrated in Fig. \ref{fig:setup1}(b). The rationale of the choice of this rheological protocol is to be able to correlate the rheological measurements with the flow measurements performed in the channel which follow a similar forcing scheme.

\begin{figure}[ht]
\centering
\subfigure[]{
     \includegraphics [width=0.4\textwidth] {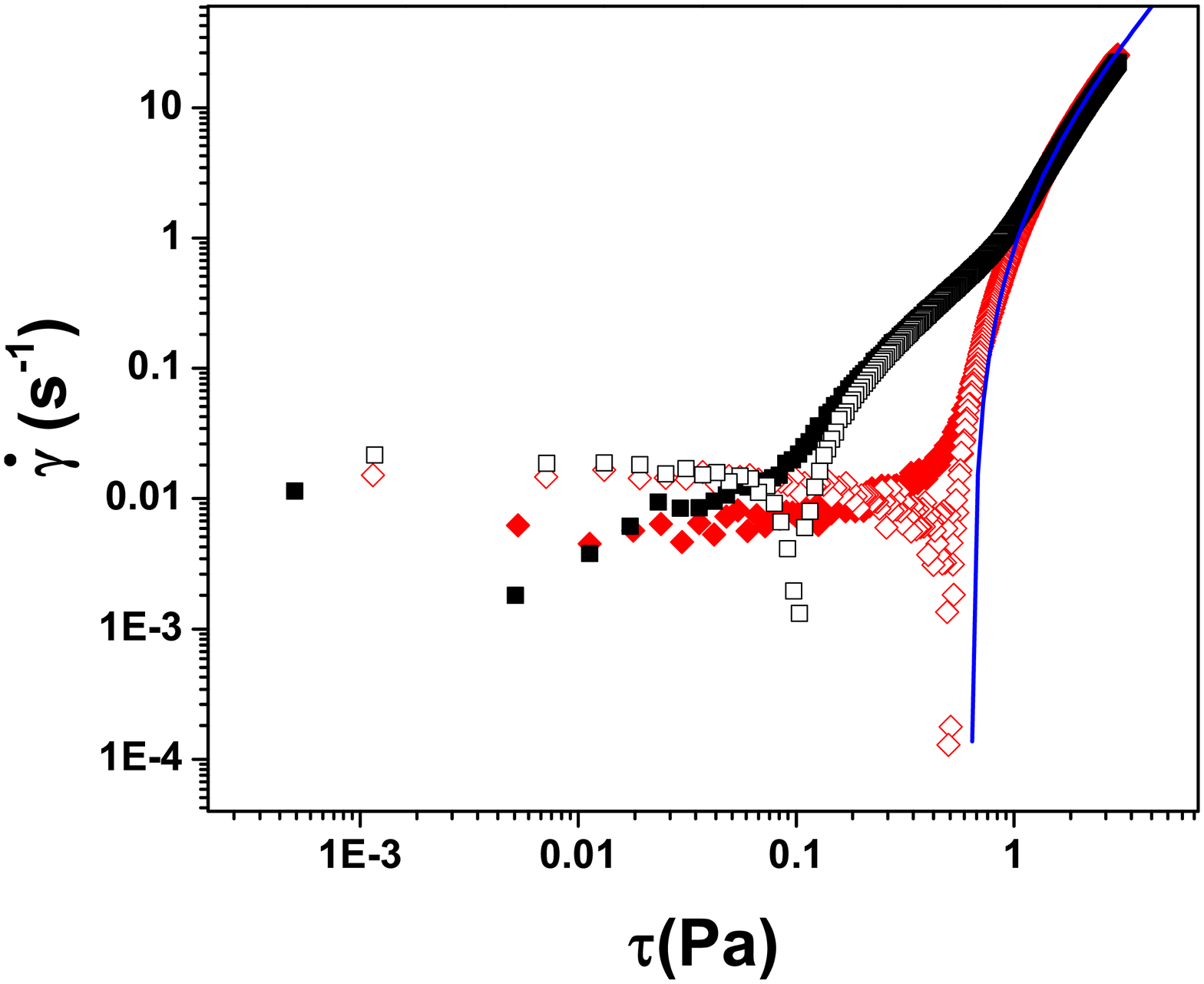}
     \label{fig:flowcurves}
}
\subfigure[]{
                 \includegraphics [width=0.4\textwidth] {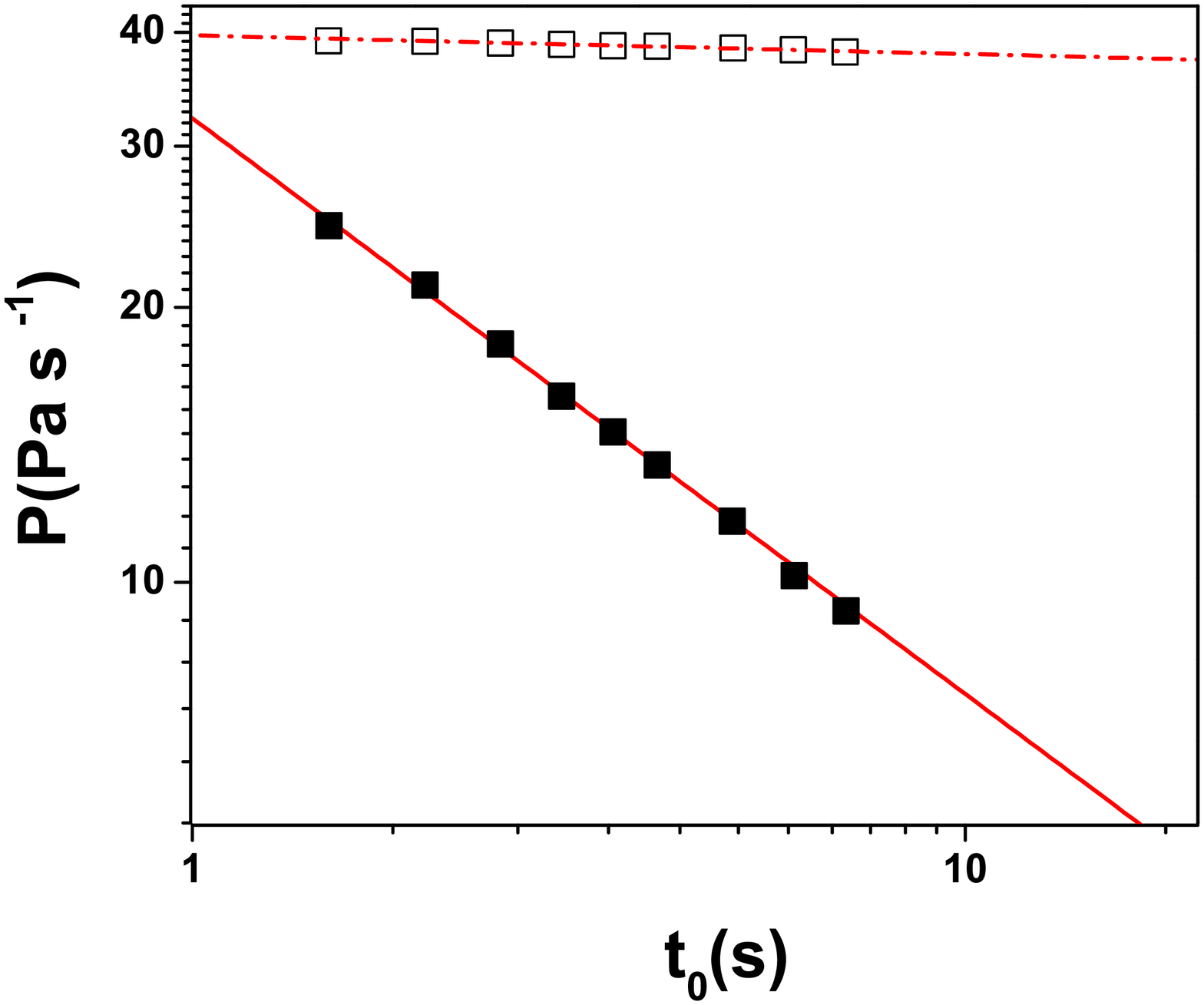}
    \label{fig:rheohysteresis}
}
\caption[Optional caption for list of figures]{\subref{fig:flowcurves} Flow curves measured with a polished geometry ($\square$, $\blacksquare$) and with a rough geometry (\textcolor{red}{$\diamond$, $\blacklozenge$}). The full/empty symbols refer to the increasing/decreasing branch of the flow curves. The full line is a Herschel-Bulkley fit that gives $\tau_y= 0.64 \pm 0.003 (Pa)$, $K= 0.4 \pm 0.002 (Pa s^n)$, $n=0.54 \pm 0.001$. \subref{fig:rheohysteresis} Dependence of the hysteresis area on the characteristic forcing time $t_0$ measured with a smooth geometry ($\square$) and a rough one ($\blacksquare$). The full line (\textcolor{red}{$-$}) is a guide for the eye, $P \propto t_0^{-0.63}$ and the dash dotted line (\textcolor{red}{$-_.-$}) is a guide for the eye, $P \propto t_0^{-0.03}$.}
\label{fig:rheology}
\end{figure}

The results of these measurements are presented in Fig. \ref{fig:rheology}. In Fig. \ref{fig:flowcurves} are displayed controlled stress ramps measured with both rough plates (the rhumbs, (\textcolor{red}{$\diamond$, $\blacklozenge$})) and smooth plates (the squares, ($\square$, $\blacksquare$)). 

As already emphasised in \cite{solidfluid, thermo}, our choice of representing the rheological data in the coordinates $\dot \gamma = \dot \gamma (\tau)$ (which might seem somewhat unusual to some) is motivated by the fact that this representation allows a more accurate identification of the various deformation regimes and a rigorous assessment of the reversibility of the deformation states.

For the case of the measurements in the absence of all slip (the rhumbs (\textcolor{red}{$\diamond$, $\blacklozenge$}) presented in Fig. \ref{fig:flowcurves}) and corresponding to low values of the applied stress ( $\tau \leq 0.64 Pa$ ) the measured rate of deformation is constant for both branches (increasing/decreasing) of the stress ramp and, consequently, the strain $\gamma$ is a linear function of the applied stresses which is the signature of a elastic solid deformation regime \footnote{Note that the stress ramp is linear in time as illustrated in Fig. \ref{fig:setup1} and, therefore,  one can conclude that the material deforms as an elastic solid within this regime, $G=\tau / \gamma=ct$}. Within this regime the deformation states are not reversible upon increasing/decreasing applied stresses which translates into different elastic moduli: $G_u>G_d$. Here the indices denote the increasing/decreasing stress ramps, respectively.

For large values of the applied stresses ( $\tau \geq 0.64 Pa$) a reversible fluid regime is observed.
One has to emphasise that the transition from a irreversible solid like deformation regime to a reversible fluid one is not direct, but mediated by an intermediate deformation regime which can not be associated with neither a solid like behaviour nor a fluid one but with a coexistence of the two phases. A similar smooth solid-fluid transition has been recently observed in a low Reynolds number steady pipe flow, \cite{pipesteady}.

The irreversibility of the deformation states observed within the solid and the intermediate deformation regimes and initially reported by Putz et al in \cite{solidfluid} has been recently confirmed by others, \cite{divoux2, divoux3, divoux4}.
A similar irreversible rheological behaviour has been found in rheological studies performed on various grades of \trademark {Carbopol} gels using various rheometers (and various geometries) at various temperatures, \cite{miguelstab, thermo}. 

We note that, during each of the rheological measurements reported in this manuscript and in our previous studies \cite{solidfluid, miguelstab, thermo}, the Reynolds number was smaller than unity, therefore the experimentally observed irreversibility of the deformation states illustrated in Fig. \ref{fig:flowcurves} should not be associated with inertial effects.

Thus, although \trademark{Carbopol} gels have been long considered as \textit{"model"} yield stress fluids that display no thixotropic effects and their rheological behaviour can be accurately described by the Herschel-Bulkley picture, this recently found irreversible rheological behaviour should be accepted as a physical fact and consequently trigger 
new developments that can properly account for this picture and its impact on the description of various realistic viscoplastic flows. 

To understand the role of wall slip in the yielding of the \trademark{Carbopol}, we have performed the same type of rheological measurements with smooth parallel plates. The main effect of the wall slip is to shift the solid-fluid coexistence regime to lower values of the applied stresses (see the squares ($\square$, $\blacksquare$) in Fig. \ref{fig:flowcurves}). The apparent yield stress measured in the presence of wall slip $\tau_y^a$ is nearly an order of magnitude smaller than the yield stress measured in the absence of slip, $\tau_y$. The data acquired with and without wall slip overlap only within the fluid regime, corresponding to shear rates $\dot \gamma >\dot \gamma_c$ (with $\dot \gamma_c \approx 0.8 s^{-1}$ see Fig. \ref{fig:flowcurves}).  This critical value of the shear rate above which the slip and no slip data overlap is comparable to the critical shear rate found by Bertola and his coworkers below which no steady state flow could be observed by MRI, \cite{bertolaslip}. The existence of the critical shear rate $\dot \gamma_c$ is also consistent with the measurements of the velocity profiles performed by Salmon and his coworkers with a concentrated emulsion sheared in a small gap Couette geometry, \cite{salmon}.
 
Additionally, we note that it is solely within the viscous deformation regime that all the data sets can be reliably fitted by the Herschel-Bulkley model (the full line in Fig. \ref{fig:flowcurves}) which reinforces the idea that only within this deformation range the \trademark{Carbopol} gel behaves like a simple or "model" yield stress fluid. 

A next fundamental question that deserves being addressed is how the irreversible character of the rheological flow curves observed within the solid-fluid coexistence regime in the form of a hysteresis loop is related to the rate at which the deformation energy is transferred to the material or, in other words, to the characteristic forcing time $t_0$. It is equally important to understand if the presence of wall slip influences this dependence. 
To address this points we compare measurements of the area of the hysteresis observed in the flow curves presented in Fig. \ref{fig:flowcurves} $P=\int \dot{\gamma}^u d \Delta \tau^u-\int \dot{\gamma}^d d \Delta \tau^d$ performed for various values of $t_0$ for both slip and no-slip cases. Here the indices "u,d" refer to the increasing/decreasing stress branches of the flow ramp. As already discussed in Ref. \cite{solidfluid} the area $P$ has the dimensions of a deformation power deficit per unit volume of sheared material. 
The results of these comparative measurements are presented in Fig. \ref{fig:rheohysteresis}.
In the absence of wall slip, the deformation power deficit scales with the characteristic forcing time as $P \propto t_0^{-0.63}$ (the full squares ($\blacksquare$) in Fig. \ref{fig:rheohysteresis}). A similar scaling has been found in the absence of wall slip for several \trademark{Carbopol} solutions with various concentrations and at various temperatures in Ref. \cite{solidfluid}, $P \propto t_0^{-1}$. The difference in the scaling exponent found in the present study and the one initially reported Ref. \cite{solidfluid} may be related to differences in the grade of the \trademark{Carbopol}.  

The existence of a rheological hysteresis for \trademark{Carbopol} gels characterised by a decrease of the power deficit with the characteristic forcing time consistent with both our initial finding reported in Ref. \cite{solidfluid} and the data presented in Fig. \ref{fig:rheohysteresis} was afterwards confirmed by others in independent experiments performed with different \trademark{Carbopol} solutions and using different experimental protocols, \cite{divoux4}.

The next question we address is what is the influence of the wall slip phenomenon on the rheological hysteresis observed during stepped stress ramps and on its scaling with the characteristic forcing time $t_0$. To address this points similar measurements of the hysteresis area $P$ for various values of $t_0$ have been performed in the presence of wall slip (the empty squares ($\square$) in Fig. \ref{fig:rheohysteresis}). The presence of the wall slip affects both the magnitude of the deformation power deficit and its scaling with the characteristic forcing time by nearly suppressing it: $P \propto t_0^{-0.03}$ (the dash-dotted line (\textcolor{red}{$-_.-$}) in Fig. \ref{fig:rheohysteresis}).

The central conclusion of this comparative investigation of the solid-fluid transition in a rheometric flow in the presence and in the absence of wall slip is two-fold. Firstly, the wall slip phenomenon and the irreversible deformation are observed in the same range of applied stresses and measured rates of shear. This means that, for this particular flow, the irreversible solid-fluid transition and the wall slip are coupled phenomena. Secondly and reinforcing the first part of the conclusion, the emergence of the wall slip modifies the irreversible solid-fluid transition by suppressing the dependence of the hysteresis power loss on the rate at which the material is being forced, $1/t_0$. 

From a theoretical point of view, the experimentally observed coupling between the wall slip and the yielding phenomena  
in a unsteady rheometric flow with slippery boundaries is still insufficiently documented and complicates the physical picture.

In the absence of wall slip, a simple model able to describe the irreversibility of deformation states observed in Fig. \ref{fig:flowcurves} has been recently proposed, \cite{solidfluid}.
As previously suggested by several authors
(\cite{moller,dullaert}), the yilding process of a physical
gel sample under shear was interpreted in terms of a dissociation
reaction, $S \rightleftharpoons S+F$ and be modelled by the
following kinetic equation:
\begin{equation}\label{kineticeq}
\frac{d
\Phi}{dt}=R_d(\Phi,t,\tau)+R_r(\Phi,t,\tau)
\end{equation}
where $S,F$ denote the solid and fluid phases, respectively,
$\Phi=[S]$ is the concentration of the solid phase,
$\tau$ is the forcing
parameter (the applied stress), $R_d=-K_d\left[ 1+\tanh\left(\frac{\tau-\tau_y}{w}\right)\right]\Phi$ is the rate of destruction of solid units and
$R_r= K_r \left [ 1- \tanh \left (
\frac{\tau-\tau_y}{w} \right ) \right ](1-\Phi)$ is the rate of fluid recombination of fluid elements into a
gelled structure, \cite{solidfluid}.

As a constitutive equation a thixo-elastic Maxwell (\textbf{TEM}) type model was used:

\begin{equation}\label{constitutive}
   \frac{\eta(\dot{\gamma})}{G} \Phi \frac{d \tau}{dt}+\tau= \eta(\dot{\gamma})\dot{\gamma}
\end{equation}
where the viscosity of the fluid phase is given by a regularized Herschel-Bulkley model, $\eta(\dot{\gamma})=K~|\dot{\gamma}|^{n-1}+\tau_y\frac{1-\exp{(-m|\dot{\gamma}|)}}{|\dot{\gamma}|}$. Here $G$ is the static elastic modulus, $K$ the
consistency, $n$ the power law index and $m$ is the
regularisation parameter.

The choice of this constitutive equation is motivated by the
presence of elastic effects in the intermediate deformation regime
(see the cusp in decreasing stress branch in Fig. \ref{fig:flowcurves} and
the corresponding discussion). We note that in the
limit $\Phi \rightarrow 1$ equation \ref{constitutive} reduces to
Hooke's law, $\tau=G\gamma$ (which describes the solid deformation regime), and in the limit $\Phi \rightarrow 0$ it reduces to a regularized Herschel-Bulkley model (which describes the viscous regime).

In spite of its simplicity, the model briefly presented above proved its ability of describing a large number of rheological flow curves similar to that presented in Fig. \ref{fig:flowcurves} (the squares ($\square$, $\blacksquare$)) measured in the absence of wall slip  for \trademark{Carbopol} solutions of various grades, concentrations and at various temperatures, \cite{solidfluid, thermo}.
Additionally and quite remarkably, for a given aqueous gel solution ($0.1 \%$, \trademark{Carbopol} 940) and for the same choice of the model parameters ($\tau_y$, $G$, $n$, $K_u$, $K_d$, $w$), the model accurately fits both controlled stress increasing/decreasing ramps and controlled stress oscillatory sweeps in a large range of oscillation amplitudes and frequencies, \cite{solidfluid}. 

The presence of wall slip which artificially diminishes the apparent yield stress $\tau_y^a<\tau_y$ and significantly alters the shape of the flow curves (see Fig.  \ref{fig:flowcurves}) renders the simple model presented above inapplicable in its present form as it contains no terms that can account for the slip behaviour and for its coupling to the solid-fluid transition. A modified version of this model that can account for wall slip by introducing a delay term in the phase equation Eq. \ref{kineticeq} will be described elsewhere.

\subsection{\label{subsec:pipeflow}The solid-fluid transition in a unsteady pipe flow}

Following the comparative investigation of the yielding of a \trademark{Carbopol} gel in a rheometric flow in the presence and in the absence of wall slip presented in Sec. \ref{sec:sfrheo}, the main question that arises is to what extend the findings on the solid-fluid transition investigated in this particular laboratory scale flow could be transferred to flows that are more relevant from a practical perspective, such as a slow (inertia-free) unsteady pipe flow. Of equal importance is understanding the role of wall slip in the dynamics of the solid-fluid transition. 

\subsubsection{\label{subsec:flow fields}Dynamics and structure of the unsteady flow fields}

Within the framework of the Herschel-Bulkley model and in the absence of wall slip, the mean flow velocity $U_{av}$ is related to the driving pressure drop via: 

\begin{equation} \label{Eq:meanflow}
  U_{av}=\begin{cases}
    0, & \text{$\xi \leq 1$}.\\
    \left [ \frac{R \Delta p}{L K} \right]^{1/n} \frac{R}{2+1/n} \left [1-\frac{1}{\xi} \right] ^{1+1/n}\left[1+ \frac{1}{\xi (1+1/n)} \right], & \text{$\xi > 1$}.
  \end{cases}
\end{equation}
see e.g. \cite{pipehb}. The dimensionless parameter $\xi$ is defined: 
\begin{equation} \label{Eq:xi}
  \xi=\frac{R \Delta p}{2 L \tau_y}
\end{equation} 
The Eq. \ref{Eq:meanflow} relies heavily on the Herschel-Bulkley yielding picture, the flow steadiness and the absence of velocity slip at the channel's wall. Although we expect that none of these assumptions are fully fulfilled by the flow we investigate, we still believe it is reasonable to use it in order to get just an indicator for order of magnitude for the applied pressure drop that corresponds to yielded states, $\Delta p_y$, which would ultimately indicate us a suitable range of the driving pressure drops $\Delta p$ that needs to be explored during our experiments in order to systematically characterise the solid-fluid transition for this type of unsteady pipe flow.

Using the yielding criterion $\xi \approx 1$ for case of the flow channel described in Sec. \ref{subsec:setup} and the \trademark{Carbopol} solution characterised in Secs. \ref{subsec:preparation}, \ref{subsec:rheological}, the pressure drop corresponding to the yielded (flowing) states can be estimated as $\Delta p_y \approx 774 Pa$.

To characterise the solid-fluid transition in the flow channel, a time series of flow fields has been acquired during a controlled increasing/decreasing pressure ramp (see Fig. \ref{fig:setup1}(b)).
This controlled pressure ramp is similar to the controlled stress ramps used to characterise the solid-fluid transition in a rheometric flow. 

\begin{figure*}
\begin{center}
\centering
\includegraphics [width=0.45\textwidth] {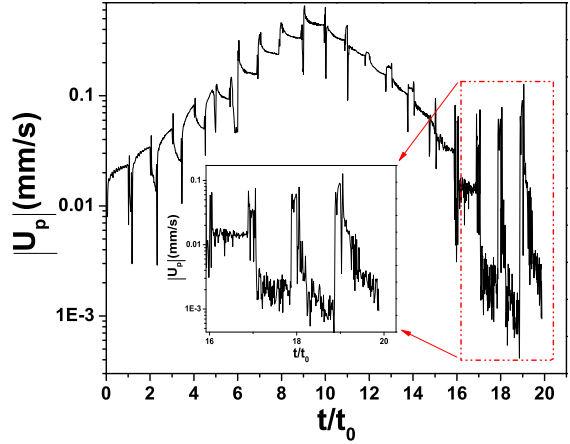}
\caption{Time series of the absolute value of the plug velocity $U_p= U\bigg |_{r= 0}$. The insert presents a magnified view of the time series after the recoil effect (see the text for the explanation) is observed. The characteristic forcing time was $t_0=7.5 s$.} \label{fig:time series}
\end{center}
\end{figure*}

The corresponding time series of the absolute value of the centre line velocity $\left| U_p \right|$ is presented in Fig. \ref{fig:time series}. The time interval during each the applied pressure was maintained constant (the characteristic forcing time $t_0$) was fixed at $7.5~s$.

In order to capture the details of the solid-fluid transition in an unsteadily forced flow, the maximal value of the applied pressure drop has been chosen nearly three times larger than the above estimated pressure drop corresponding to the yielded state, $\Delta p_y$. 

The spikes visible in the velocity time series presented in Fig. \ref{fig:time series} correspond to the change of the driving pressure realised by the displacement of the vertical translational stage (\textbf{TS} in Fig. \ref{fig:setup1}(a)) which carries the inlet fluid container \textbf{IFC}. Its magnitude depends on the inertia of the mechanical system and the acceleration of the stepping motor \textbf{M}. 

To avoid accounting for this response of the mechanical system which controls the pressure drop along the flow channel, the flow data acquired during these transients (extended over roughly $0.5s$ in each of the experiments reported here) has been discarded.     

Corresponding to the first five values of the driving pressure on the increasing branch of the ramp the time dependence of the centreline velocity does not reach a steady state during the characteristic forcing time $t_0=7.5 s$.
 
A further increase of the driving pressure reveals an interesting feature of the velocity time series in the form of a non-monotone time dependence: upon the change of the pressure drop, the centreline velocity first decreases and then increases. A phenomenological explanation of this observation may be given in terms of a strong heterogeneity of the flow, i.e. a dynamic coexistence and competition between yielding of solid material elements and the re-combination of yielded fluid elements which, as illustrated by the rheometric data presented in Fig. \ref{fig:flowcurves}, occurs within an intermediate range of the applied stresses (pressure drops). Indeed, as the driving pressure drops $\Delta p$ are further increased, this non-monotone behaviour is no longer observed and the plug velocity reaches a steady state during a characteristic time smaller than $t_0$. This indicates that, within this range of driving pressure, a steady yielded state is achieved, Fig. \ref{fig:time series}.

The time series presented in Fig. \ref{fig:time series} may also provide a first indication on the reversibility of the flow states upon increasing/decreasing the driving pressure drops. Thus, one can note that the data is not symmetric with respect to the vertical line $t/t_0=10$. This observation is consistent with the irreversibility of the deformation states observed for the rheological data acquired at low and intermediate values of the applied stresses, Fig.\ref{fig:flowcurves}. An even clearer signature of the irreversibility of the deformation states can be noted if one monitors the time series data acquired during the last four steps of the decreasing branch of the pressure ramp and highlighted in the insert of Fig. \ref{fig:time series}. For these steps of the pressure ramp velocity fluctuates strongly and, corresponding to the the last step a flow reversal is apparent: the absolute value of velocity seems to increase as the pressure drop is decreased to its smallest value. A careful inspection of the flow images acquired during this step clearly indicate a reversal of the flow direction. This rather unexpected effect can be associated to a elastic recoil effect similar to the recoil observed during the rheological measurements on the decreasing stress branch and manifested by the cusp visible in Fig.  \ref{fig:flowcurves}.

\begin{figure*}
\begin{center}
\centering
\includegraphics [width=0.8\textwidth] {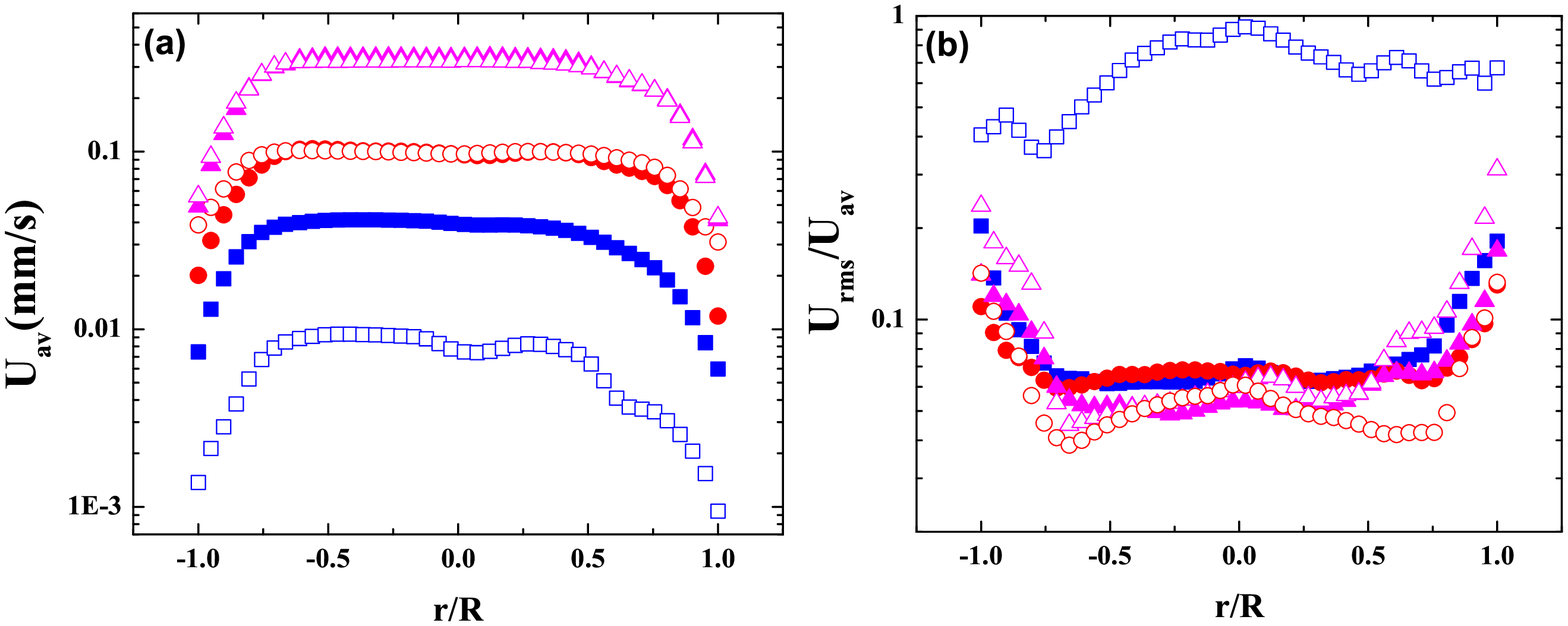}
\caption{(a) Profiles of the absolute value of the mean flow velocity $U_{av}$. (b) Profiles of the reduced root mean square variation of the velocity, $U_{rms}/U_{av}$.  The symbols refer to different values of the pressure drop: ($\textcolor{blue}{\square, \blacksquare}$) - $\Delta p=310.8 Pa$,  ($\textcolor{red}{\circ, \bullet}$) - $\Delta p=621.6 Pa$,   ($\textcolor{magenta}{\bigtriangleup, \blacktriangle}$) - $\Delta p=932.3 Pa$ . The full/empty symbols refer to the data acquired on the increasing/decreasing branch of the pressure ramp, respectively. The characteristic forcing time is $t_0=10 s$.} \label{fig:profiles}
\end{center}
\end{figure*}

We focus in the following on the structure of the unsteady flow fields and their fluctuations.  
Profiles of the absolute value of the mean flow velocity $U_{av}$ and the reduced velocity fluctuation $U_{rms}/U_{av}$ measured for several values of the driving pressure $\Delta p$ on both the increasing and the decreasing branch of the flow ramp illustrated in Fig. \ref{fig:setup1}(b)  are presented in Fig. \ref{fig:profiles}.

For each value of the driving pressure, on the increasing branch of the flow ramp the mean velocity profile indicates the presence of a central unyielded plug, Fig. \ref{fig:profiles} as typically expected for a viscoplastic fluid. Also, for each value of the pressure drop, a slip behaviour is observed near the channel walls, $r= \pm 1$. For the largest values of the driving pressure ($\Delta p = 932.3 Pa$ and $\Delta p= 621.6 Pa$, (the triangles ($\textcolor{magenta}{\bigtriangleup, \blacktriangle}$) in Fig. \ref{fig:profiles} (a))) the velocity profiles measured on the increasing/decreasing branches of the ramp coincide consistently with the time series measurements presented in Fig. \ref{fig:time series}. 
 This indicates that a stationary fluid flow regime has been achieved. Within this state the velocity fluctuations quantified by $U_{rms}/U_{av}$ are the smallest and mainly localised near the channel wall which is another indicator for the steadiness of the flow (see Fig. \ref{fig:profiles} (b)). 
 
Corresponding to the lowest driving pressure, however, the profile measured on the decreasing branch (the empty squares, $\textcolor{blue}{\square}$, in Fig. \ref{fig:profiles} (a)) falls clearly below the profile measured on the increasing pressure branch. This indicates that within this range of pressure drops the flow is irreversible upon increasing/decreasing the forcing. 
It is equally worth noting that, corresponding to this driving pressure, the mean flow profile is less smooth around the centreline of the channel which indicates a rather large level of velocity fluctuations. It also indicates that smaller the driving pressure is further the flow is from a steady state. 

To probe this hypothesis we turn our attention to the transversal profiles of the reduced velocity fluctuations $U_{rms}/U_{av}$, Fig. \ref{fig:profiles} (b)). For large driving pressure (the triangles in Fig. \ref{fig:profiles} (b)) the reduced velocity variations account for roughly $6 \%$ of the mean flow velocity in the bulk region of the flow (around the unyielded plug) and they may reach up to $20 \%$ of the mean flow velocity in the wall region (due to the smallness of the measured velocity).
The rather large values of the fluctuations observed near the channel walls are instrumental and they are due to the smallness of the measured velocity within this regions (here, the \textbf{DPIV} heavily relies on the sub-pixel interpolation because of the very small displacements of the flow tracers). 

 On the decreasing branch of the controlled pressure ramp and for low driving pressures a localisation of the velocity fluctuations can be observed in the form of a local maximum centred around the middle of the channel ($r/R=0$) can be observed (the empty circles ($\textcolor{red}{\circ}$), the empty triangles ($\textcolor{magenta}{\bigtriangleup}$) and the empty squares ($\textcolor{blue}{\square}$) in Fig. \ref{fig:profiles}(b)). The localisation of the velocity fluctuations near the centre line of the flow channel may be understood in terms of large fluctuations of a elastic (un-yielded) plug which is consistent with the experimental findings for the case of a oscillatory pipe flow, \cite{park}.

\subsubsection{\label{subsec:reversibility}Reversibility of the flow states, the role of wall slip}

We focus in the following on the reversibility of the flow states upon increasing/decreasing pressure drop and its connection to the irreversible solid-fluid transition observed in a rheometric flow in the presence of wall slip illustrated in Fig. \ref{fig:rheology}.
 
We note that our flow investigation technique required us to use a flow channel with slippery walls so a comparison of the slip and no slip cases cannot be performed for this pressure driven flow and, consequently, the yielding transition can not be decoupled from the wall slip. 

Such a comparative study remains open to future experimental investigations  
that use measuring techniques that do not require optical transparency of the flow channels such as ultrasonic velocimetry. 

\begin{figure*}
\begin{center}
\centering
\includegraphics [width=0.6\textwidth] {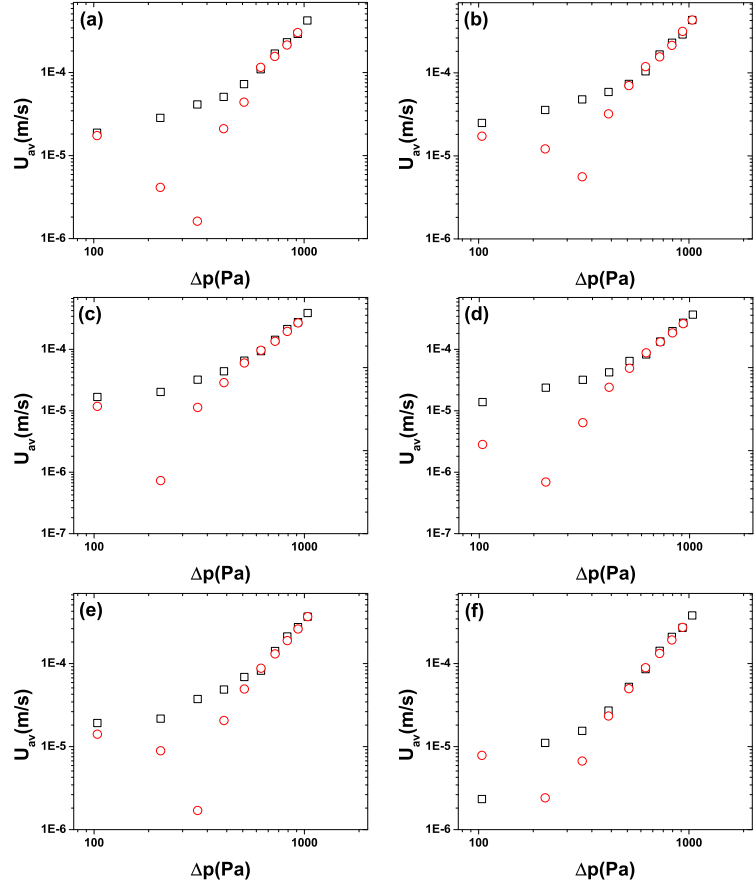}
\caption{Dependence of the absolute value of the mean flow velocity $U_{av}$ on the applied pressure drop $\Delta p$ for various values of the characteristic forcing time: (a) $t_0=4 s$, (b) $t_0=5 s$, (c) $t_0=7.5 s$, (d) $t_0=10 s$, (e) $t_0=15 s$, (b) $t_0=20 s$. The squares ($\textcolor{black}{\square}$) refer to the increasing branch of the stress ramp and the circles ($\textcolor{red}{\circ}$) refer to the decreasing branch of the stress ramp.
} \label{fig:sspeedhysteresis}
\end{center}
\end{figure*}

Measurements of the absolute value of the mean flow velocity $U_{av}$ performed for increasing/decreasing values of the driving pressure $\Delta p$ and several values of the characteristic forcing time are presented in Fig. \ref{fig:sspeedhysteresis}. The mean flow velocity $U_{av}$ has been calculated by numerical integration of the \textbf{DPIV} measured velocity profiles, $U_{av} = 2\pi\int_0^R r U(r) dr$ \footnote{The axial symmetry of the flow has been assumed.}. 

Regardless the value of the characteristic forcing time, the data acquired for increasing/decreasing driving pressures overlap only within the fluid regime $\Delta p> \Delta p_y$. This indicates that only corresponding to this range of the driving pressures the \trademark{Carbopol} gel is fully yielded. Below this critical value of the driving pressure a hysteresis of the flow states is observed. This observation is fully consistent with the rheological hysteresis observed in the presence of wall slip (the squares ($\textcolor{black}{\square}$) in Fig. \ref{fig:flowcurves}). As previously mentioned through our paper, the cusp visible on the decreasing pressure branch corresponds to a elastic recoil effect manifested by a reversal of the flow direction. This indicates that, as in the case of the solid-fluid transition observed in a rheometric flow \cite{solidfluid},   the elasticity cannot be ignored while studying the unsteady yielding in a slow pipe flow flow of a \trademark{Carbopol} gel.     
 
 \begin{figure*}
\begin{center}
\centering
\includegraphics [width=0.4\textwidth] {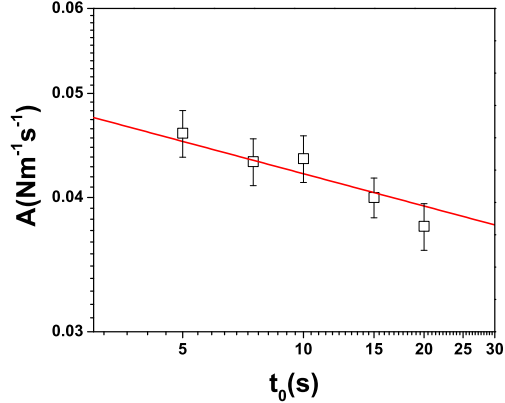}
\caption{ Dependence of the area of the hysteresis of the dependencies of the mean flow velocity on the driving pressure drop presented in Fig. \ref{fig:sspeedhysteresis} on the characteristic forcing time $t_0$.  
The full line is a guide for the eye, $A \propto t_{0}^{-0.1}$.} \label{fig:velhysteresisarea}
\end{center}
\end{figure*}

 The dependence of the hysteresis area of the dependence of the mean flow velocity $U_{av}$ on the driving pressure $\Delta p$ defined by $A=\int U_{av}^u d \Delta p^u-\int U_{av}^d d \Delta p^d$ is illustrated in Fig. \ref{fig:sspeedhysteresis} is presented in Fig. \ref{fig:velhysteresisarea}. Here the indices "u, d" refer to the increasing/ decreasing branch of the controlled pressure ramp sketched in Fig. \ref{fig:setup1} (b). As in the case of the rheological hysteresis in the presence of slip illustrated in Fig. \ref{fig:rheohysteresis} a weak dependence of the hysteresis area on the characteristic forcing time is observed, $A \propto t_0^{-0.1}$. 
 
 This fact indicates that the presence of wall slip modifies the irreversibility of the flow curves within the solid-fluid coexistence regime and prompts us to focus closer on the slip behaviour observed during the stepped pressure ramp and its relation to the rheological behaviour illustrated in Fig. \ref{fig:rheology}.

\begin{figure*}
\begin{center}
\centering
\includegraphics [width=0.6\textwidth] {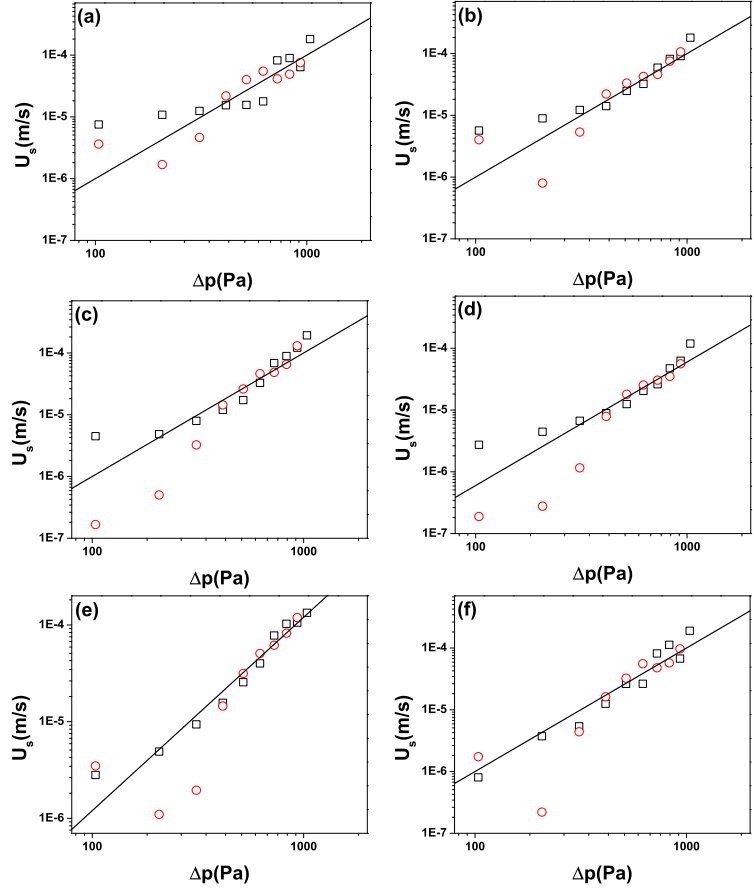}
\caption{Dependence of the slip velocity $U_{s}$ on the applied pressure drop $\Delta p$ for various values of the characteristic forcing time: (a) $t_0=4 s$, (b) $t_0=5 s$, (c) $t_0=7.5 s$, (d) $t_0=10 s$, (e) $t_0=15 s$, (b) $t_0=20 s$. The squares ($\textcolor{black}{\square}$) refer to the increasing branch of the stress ramp and the circles ($\textcolor{red}{\circ}$) refer to the decreasing branch of the stress ramp. The full lines in each panel are guides for the eye, $U_s \propto \Delta p ^2$.
} \label{fig:sliphysteresis}
\end{center}
\end{figure*}

The dependence of the slip velocity $U_s$ on the driving pressure drop $\Delta p$ measured for several values of the characteristic forcing time $t_0$ is presented in Fig. \ref{fig:sliphysteresis}. 

As for the pressure dependence of the mean flow velocity $U_{av}$ illustrated in Fig. \ref{fig:sspeedhysteresis} and discussed above, one can clearly see in Fig. \ref{fig:sliphysteresis} that, regardless the value of $t_0$, the dependence of the slip velocity on the applied pressure drop $\Delta p$ is not reversible upon increasing/decreasing pressure over the entire range of pressures. This is another indicator that the slip behaviour can not be entirely decoupled from the irreversible solid-fluid transition observed during the rheological measurements presented in Fig. \ref{fig:flowcurves}. Regardless the value of the characteristic forcing time $t_0$, within the fluid flow regime ($\Delta p > 500 Pa$) the slip velocity scales with the applied pressure drop as $U_s \propto \Delta p ^2$ (see the full lines in Fig. \ref{fig:sliphysteresis}). This scaling law differs from the one found by Gonz\'alez and his coworkers in \cite{pipesteady}, $U_s \propto \Delta p^{0.876}$ (see Fig. 6 in Ref. \cite{pipesteady} in connection to their Eq. (3)). As the scaling law we found is practically insensitive to the characteristic forcing time, we believe that this discrepancy does not originate in the steadiness of the pipe flow investigated in \cite{pipesteady} but in the different rheological properties of the \trademark{Carbopol} solutions they have used (nearly an order of magnitude difference in both the yield stress and the consistency). Thus, from this comparison, it is apparent that even within the reversible fluid regime ($\tau>\tau_y$) the wall slip behaviour remains correlated to the rheological properties of the solution.  

\begin{figure} 
\begin{center}
\centering
  \includegraphics [width=0.4\textwidth] {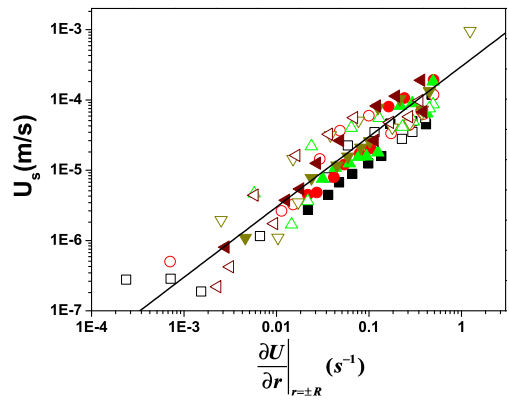}
\caption{Dependence of the slip velocity $U_s$ on the wall velocity gradients. The full line is a guide for the eye, $U_s \approx 3 \cdot 10^{-4} \frac{\partial U}{\partial r} \bigg |_{r=\pm R} $. The symbols are: (\textcolor{green}{$\bigtriangleup$, $\blacktriangle$}) - $t_0=4~(s)$,  (\textcolor{red}{$\circ$, $\bullet$}) - $t_0=7.5~(s)$,
 (\textcolor{black}{$\square$, $\blacksquare$}) - $t_0=10~(s)$, (\textcolor{olive}{$\triangledown$, $\blacktriangle$}) - $t_0=15~(s)$, (\textcolor{brown}{$\triangleleft$, $\blacktriangleleft$}) - $t_0=20~(s)$.}\label{fig:slipvsgrad}
\end{center}
\end{figure}

To get a deeper insight into this correlation, we focus on the dependence of the slip velocity $U_s$ on the wall velocity gradients $\frac{\partial U}{\partial r} \bigg |_{r=\pm R} $, Fig.  \ref{fig:slipvsgrad}. The measurements presented in Fig.  \ref{fig:slipvsgrad} are performed for various values of the characteristic forcing time $t_0$ on both the increasing and the decreasing branch of the pressure ramp (see Fig. \ref{fig:setup1} (b)).

For each of the data sets presented, the wall velocity gradients have been obtained by numerical differentiation of the \textbf{DPIV} measured averaged velocity profiles near the wall and the slip velocity has been obtained by extrapolating the same time averaged profiles at the wall (see Fig. \ref{fig:profiles} (a)). To increase the accuracy of the numerical differentiation, a central difference differentiation scheme has been used. 

Previous works have reported a decrease of the slip velocity with the applied stresses (velocity gradients) beyond the yield-point, \cite{bogerreview}. Contrarily to this, our experimental findings indicate a monotone increase of the slip velocity with the wall velocity gradients, regardless the deformation regime (solid, fluid or solid-fluid) and the degree of flow steadiness (the value of $t_0$). This conclusion is consistent, however, with the experimental findings for the case of a steady pipe flow, \cite{pipesteady}. 

\begin{figure} 
\begin{center}
\centering
  \includegraphics [width=0.4\textwidth] {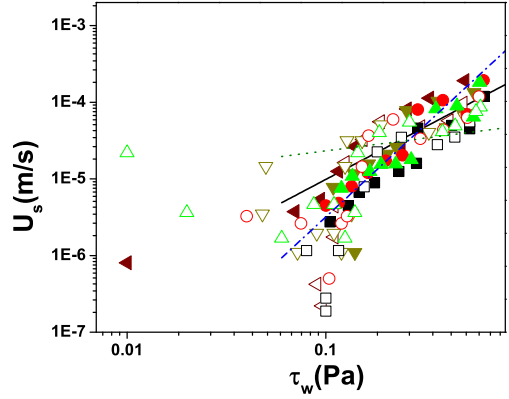}
\caption{Dependence of the slip velocity $U_s$ on the wall shear stresses $\tau_w$. The full line is a power law fit, $U_s=2.3 \cdot10^{-4} (\pm 5.4 \cdot 10^{-5}) \cdot \tau_w^{1.32 \pm 0.26}$. The dotted and the dash-dotted lines are the theoretical predictions by Piau (see Ref. \cite{piaucarbopol}): $U_s \propto \tau_w^{1/3}$, $U_s \propto \tau_w^{2}$. The symbols are: (\textcolor{green}{$\bigtriangleup$, $\blacktriangle$}) - $t_0=4~(s)$,  (\textcolor{red}{$\circ$, $\bullet$}) - $t_0=7.5~(s)$,
 (\textcolor{black}{$\square$, $\blacksquare$}) - $t_0=10~(s)$, (\textcolor{olive}{$\triangledown$, $\blacktriangle$}) - $t_0=15~(s)$, (\textcolor{brown}{$\triangleleft$, $\blacktriangleleft$}) - $t_0=20~(s)$.}\label{fig:slipvsstresses}
 \end{center}
\end{figure}

Regardless the deformation regime, the value of the characteristic forcing time $t_0$ and the type of the pressure ramp (increasing or decreasing pressures) a universal dependence of the slip velocity $U_s$ on the wall velocity gradients $\frac{\partial U}{\partial r} \bigg |_{r=\pm R}$ is observed, Fig. \ref{fig:slipvsgrad}, and all the data follow a line who's slope indicates a slip length of roughly $300 \mu m$ (the full line in Fig. \ref{fig:slipvsgrad})

 Thus, one can conclude from theses measurements that the dependence of the slip velocity on the wall velocity gradients has an universal character and uniquely defines a slip length which is independent on the deformation regime but may depend on the roughness of the channel wall and on the physicochemical properties of the \trademark{Carbopol} solution. Two natural but fundamental questions arise at this point: 
\begin{enumerate}
\item \textit{What is the extent of this universal behaviour?} 
\item \textit{Is there a simple phenomenological picture underlying the wall slip behaviour observed for a \trademark{Carbopol} gel?}
\end{enumerate}

A first phenomenological picture for the wall slip phenomenon (observed in suspensions) has been suggested in early $20's$ by Bingham by the following quote: \textit{"Slip comes from a lack of adhesion between the material and the shearing surface. The result is that there is a layer of liquid between the shearing surface and the main body of the suspension"}, see page $231$ in Ref. \cite{binghambook}.

Kaylon developed further on the slip picture suggested by Bingham and proposed the following relationship between the slip velocity and the wall shear stress, \cite{kaylonslip}:

\begin{equation} \label{eq:slip}
U_s=\beta \left ( \tau_w \right )^{1/n_s}
\end{equation}
The slip coefficient $\beta$ is related to the width $\delta$ of the slip layer,  the consistency $K_s$ and power law index $n_s$ of the interstitial fluid arrested within the slip layer (referred to in Ref. \cite{kaylonslip} as the \textit{"binder fluid"}) via $\beta=\frac{\delta}{K_s^{n_s}}$.

\trademark{Carbopol} micro gels are not suspensions in the strict sense of the term but they can still be viewed as a jammed system of swollen gel micro-particles, \cite{piaucarbopol}. Thus, bearing in mind that the gel was prepared in an aqueous solvent, it looks reasonable to us to consider the \textit{"binder fluid"} as Newtonian, $K_s=1~mPas, n_s=1$.   
 
To test the applicability of Eq. \ref{eq:slip} with the assumptions above for our data, we re-plot the slip velocity data presented in Fig. \ref{fig:slipvsgrad} versus the wall shear stress defined by $\tau_w=\frac{\partial U}{\partial r} \bigg |_{r=\pm R} \cdot \eta \left ({\frac{\partial U}{\partial r} \bigg |_{r=\pm R}} \right )$. For this purpose we have related the wall velocity gradients measured by numerical differentiation of the \textbf{DPIV} measured velocity profiles to the wall shear stresses via the rheological flow curve measured in the presence of wall slip and presented in Fig. \ref{fig:flowcurves} (the squares) with the wall velocity gradient playing the role of the rate of shear $\dot \gamma$. The result of this evaluation is presented in Fig. \ref{fig:slipvsstresses}. Within this representation the universality observed in Fig. \ref{fig:slipvsgrad} is broken and three different slip regimes can now be clearly observed. Corresponding to values of the wall shear stress significantly smaller then the apparent yield stress measured in the presence of wall slip ($\tau_w < \tau^a_y \approx 0.1 Pa$, see Fig. \ref{fig:flowcurves}) the measured slip velocity displays a rather large scatter, most probably related to the smallness of both the slip velocity and wall velocity gradients. However, in spite of this data scatter  and as wall shear stress approaches the apparent yield stress $\tau^a_y$ measured in the presence of wall slip, one can note a non-monotone dependence of the slip velocity on the wall shear stress which is a reminiscent of the cusp observed in the dependence of the rate of shear on the applied stress presented in Fig. \ref{fig:flowcurves}. This behaviour clearly departs from a Navier-like slip behaviour and it is related to the non-trivial yielding process illustrated in Fig. \ref{fig:flowcurves} which occurs via intermediate range of stresses where the solid and fluid phases coexist, \cite{solidfluid}. Above the apparent yield stress and corresponding to the fluid deformation regime a nearly linear scaling of the slip velocity with the wall shear stress is observed, $U_s \propto \tau_w^{1.32\pm0.26}$ (the full line in Fig. \ref{fig:slipvsstresses}). This indicates a Navier-like slip behaviour which can be described by Eq. \ref{eq:slip}. This nearly linear scaling reinforces the assumption above that the \textit{"binder fluid"} is Netwonian, i.e. $n_s \approx 1$. We note that a similar scaling law has been found in the recent micro-flow experiments by Gonzalez and his coworkers, \cite{pipesteady} (see Fig. 6 in their paper):  $U_s \propto \tau_w^{0.876}$.
The pre-factor of this linear dependence is $\beta \approx 0.00023~mPa^{-1} s^{-1}$ which leads to an estimate for the thickness of the Newtonian shear layer $\delta \approx 0.23 \mu m$.  This estimate is fairly close to the value measured by Jiang and his coworkers for the case of hydroxypropyl guar (HPG) gel, $\delta \approx 0.1 \mu m$. 

It is interesting to compare our experimentally found scaling of the slip velocity with the theoretical prediction of Piau for a \trademark{Carbopol} gel presented in Ref \cite{piaucarbopol}.
Using a Hertzian framework to model the contacts between neighbouring micro-gel elastic particles (a detailed description of the theoretical framework can be found in Ref. \cite{johnsonbook}), Piau predicts two scaling laws of the slip velocity with the wall shear stress depending on the packing of the soft particles: $U_s \propto \tau_w^2$ for a closely packed system and $U_s \propto \tau_w^{1/3}$ in the case of a loosely packed system. The scaling exponent obtained by fitting the data presented in Fig. \ref{fig:slipvsstresses} corresponding to the fully yielded regime differs from these theoretical predictions (see the dotted and the dash-dotted lines in Fig. \ref{fig:slipvsstresses}). 

A possible reason of this discrepancy might originate in the fact that the interstitial lubricating fluid considered in Ref. \cite{piaucarbopol} was non-Newtonian and, to get a deeper insight into the slip behaviour of \trademark{Carbopol} gels a additional discussion of this issue is in order. A physical rationale for the choice of a non-Newtonian (shear thinning) interstitial fluid in Ref. \cite{piaucarbopol} might be to decouple at a micro scale the shear thinning behaviour (by attributing it to the interstitial fluid) from the yield stress behaviour by implicitly attributing it to the elastic-solid \trademark{Carbopol} micro particles. 

The disagreement we pointed out above seems to indicate that this decoupling is not as straightforward and should be more cautiously considered in future attempts to theoretically describe the relation between flow properties, wall slip and microstructure. Future theoretical approaches should definitely reconsider this problem.   

A more elaborated theory of the slip dynamics observed in a soft gel was  presented by Meeker and his coworkers, \cite{cloitreslip}. Due to the non-negligible level of scatter of the data presented in Fig. \ref{fig:slipvsstresses} induced by the numerical differentiation of the measured velocity profiles, however, a direct comparison of our results with these predictions is not possible.  

Currently, there exist no experimental studies of the microscopic scale dynamics of \trademark{Carbopol} gels under shear in the vicinity of slippery surfaces able to confirm the existence of a shear lubricating layer and provide a quantitative measure of its thickness. 
Thus, the estimate of the width $\delta$ of the lubrication layer we have obtained indirectly from the analysis slip velocity data should only be regarded as plausible. A direct experimental investigation by fluorescent visualization of the gel network as recently performed by Gutowski et al \cite{debruynscaling} would be highly needed to test this estimates.
We can, however, provide some additional arguments for the meaningfulness of this estimate as follows.  

Systematic microfluidic investigations of the diffusion of nano-tracers within a \trademark{Carbopol} micro-gel indicate that the average characteristic size of gel micro particles is of the order of $R_{av}\approx 1 \mu m$,  \cite{debruyn1, debruyn2}. A width of the shear layer $\delta \approx 0.23 R_{av}$ together with the assumption of a random close packing of the micro-gel particles is consistent with a rough standard deviation of the particle size $R_{std}/R_{av} \approx 23 \%$. Again, based on the microscopic experimental evidence presented in Refs. \cite{debruyn1, debruyn2} such a value can be considered, at least for now, plausible.

To conclude this part, our experimental data seems to confirm that in a fluid state ($\tau>\tau_y^a$) the Bingham inspired phenomenological picture of a slip layer proposed by Kaylon is applicable to the unsteady pipe flow of the \trademark{Carbopol} gel investigated through our paper.  

Another important fact that needs to be pointed out regarding the data presented in Fig. \ref{fig:slipvsstresses} is that, regardless the value of the characteristic forcing time $t_0$ and for both increasing/decreasing values of the driving pressure $\Delta p$ (the full/empty symbols in Fig. \ref{fig:slipvsstresses}, respectively), the slip velocity data overlap onto a single master curve. This invariance of the slip velocity data with the characteristic forcing time $t_0$ can be explained by the weak dependence  on $t_0$  of the rheological flow curves measured in the presence of wall slip illustrated in Fig. \ref{fig:rheohysteresis} which consequently translates into a weak dependence of the hysteresis of the measured mean flow speed illustrated on $t_0$ (see Fig.\ref{fig:velhysteresisarea}).

The central conclusion of the investigation of the wall slip behaviour is that, in the case of a unsteady low Reynolds number flow of a \trademark{Carbopol} solution, the slip behaviour can not be decoupled from solid-fluid transition. Future theoretical developments should account for this experimental fact.

\section{\label{sec:conclusions}Conclusions, outlook}

An experimental investigation of the slow yielding of a \trademark{Carbopol} gel in both a unsteady rheometric flow between parallel and in a inertia-free unsteady pipe flow is presented.

The first part of our study presented in Sec. \ref{sec:sfrheo} dealt with a detailed experimental investigation of the solid-fluid transition of a \trademark{Carbopol} gel observed during unsteady flow ramps performed in a parallel plate geometry mounted on the rotational rheometer.

The comparative investigations of the solid-fluid transition in the parallel plate geometry with and without slip performed for a wide range of characteristic forcing times $t_0$ lead to the following conclusions:

\begin{enumerate}
\item {In the presence of wall slip, the solid-fluid transition occurs earlier, i. e. the apparent yield stress measured in the presence of slip $\tau_y^a$ is nearly an order of magnitude smaller than the yield stress measured in the absence of wall slip, $\tau_y$ (see Fig. \ref{fig:flowcurves}).}

\item{In both the slip and the no slip case the transition from a solid behaviour to a viscous one (fully yielded)  is not direct but mediated by an intermediate solid-fluid coexistence regime. The cusp visible within the solid-fluid coexistence regime on the decreasing branch of the stress ramp is associated to a elastic-recoil effect manifested by a flow reversal.}
\item{In both the slip and the no slip cases the solid-fluid transition is not reversible upon increasing/decreasing of the applied stresses and a hysteresis is clearly visible (see Fig. \ref{fig:flowcurves}).}
\item{The presence of wall slip nearly suppresses the scaling of the deformation power deficit $P$ with the rate at which the material is forced, $1/t_0$.}
\item{The irreversible solid-fluid transition and the wall slip are observed within the same range of applied stresses indicating that the two phenomena are coupled.}

\end{enumerate}
 
 The second part of our study presented in Sec. \ref{subsec:pipeflow} concerns with a systematic experimental investigation of low Reynolds number unsteady pipe flows of a viscoplastic fluid (\trademark{Carbopol}-980). 
By means of the \textbf{DPIV} technique a full characterisation of the flow fields during controlled stepped ramps of the driving pressure (Fig. \ref{fig:setup1} (b)) is made. 
Depending on the value of the applied pressure several flow regimes are observed. 

Corresponding to the lowest values of the driving pressure $\Delta p$ a solid like flow regime is observed. This regime is characterised by a an elastic solid plug which never reaches a steady flow state. The largest level of fluctuations of the solid plug is observed near the centre line of the flow channel which we interpret as the signature of an unsteady solid (elastic) plug.

For the largest values of the driving pressure $\Delta p$ a fluid regime is observed. Within this regime the velocity reaches a steady state (see Fig. \ref{fig:time series}) and the flow is reversible upon increasing/decreasing pressures, Figs. \ref{fig:sspeedhysteresis}, \ref{fig:sliphysteresis}.
As in the case of the rheometric flow discussed in Sec. \ref{sec:sfrheo}, the transition between the solid-like regime to the fully yielded regime is not direct but mediated by an intermediate solid-fluid coexistence regime. For each value of the characteristic forcing time $t_0$ we have explored, this intermediate phase coexistence regime manifests itself by a clear hysteresis (observed upon increasing/decreasing the driving pressure drop) which is qualitatively similar to the hysteresis observed in the rheological flow curves. The experimentally observed hysteresis of the flow curves is associated with a deformation power deficit and measurements of this quantity for various values of $t_0$ reveal no sensitive dependence on the frequency of forcing, $A \propto t_0 ^{-0.1}$ (see Fig. \ref{fig:velhysteresisarea}). This fact is fully consistent with the measurements of the deformation power deficit performed in a rheometric (plate-plate) setup in the presence of slip, $P \propto t_0^{-0.03}$, Fig. \ref{fig:rheohysteresis}. 

As a conclusion for this part, the main features of the solid-fluid transition observed in a rheometric flow in the presence of wall slip are equally observed in a unsteady pipe flow. 

In spite of its ubiquitous nature and tremendous importance, the phenomenon of wall slip in pasty materials and dispersions is still under-looked by many researchers, \cite{slipletter}. As \trademark{Carbopol} gels are concerned, a detail characterisation of their slip properties in slow pipe flows was, to our best knowledge, insufficiently documented. The time and space resolved measurements of the flow fields based on the \textbf{DPIV} technique allowed us to attempt to fill this gap and provide a detailed experimental characterisation of the wall slip phenomenon.  
Systematic measurements of the slip velocity during controlled pressure ramps with various degrees of steadiness (i.e. for several values of the characterising forcing time $t_0$) reveal a clear coupling between the irreversible solid-fluid transition and the slip phenomenon. A linear scaling of the slip velocity with the wall velocity gradients is found, Fig. \ref{fig:slipvsgrad}. This scaling law is universal in the sense that it does not depend on the flow steadiness and invariantly holds for both the increasing and the decreasing branches of the controlled pressure ramp. 

A clear picture of the coupling between the irreversible solid-fluid transition and the slip phenomenon can be obtained by plotting the slip velocity against the wall shear stress, Fig. \ref{fig:slipvsstresses}. It is only within the fully yielded regime ($\Delta p>\Delta p_y$) that a Navier-type nearly linear scaling of the slip velocity can be observed. The measurement of the slip coefficient allowed one to estimate the thickness of the lubrication layer responsible for the slip behaviour. The result of this estimate seems to be consistent with previous experimental findings. It is equally interesting to note that this scaling of the slip velocity is practically independent on the flow steadiness (the characteristic forcing time $t_0$) and whether the driving pressure is increasing or decreasing.  

Beyond the main findings briefly highlighted above, the present study also brings a modest contribution from a technical point of view: our approach of correlating \textbf{DPIV} measured flow fields with rheometric flow curve data allows a local measurement of the stresses which, in other words, \textit{"converts"} the pipe flow into a table top open flow rheometer. To our best knowledge, this experimental approach is rather new (particularly in the context of viscoplastic flows) and deserves being pursued.   

We believe that our experimental findings could trigger several interesting developments at both experimental and theoretical level. 
A micro flow investigation of the wall slip of \trademark{Carbopol} gels during unsteady flows over smooth solid surfaces able to simultaneously monitor the flow fields and the micro-gel structure near the solid surface would bring in our opinion a ultimate and direct clarification on the coupling between yielding and slip as well as on the microscopic scale mechanism of slip. Though we are aware that such experiments represent a daunting task, we note that several steps have already been taken in this direction by combining microfluidics with various micro-rheology approaches (diffusion of nano tracers, \cite{debruyn1, debruyn2}, fluorescent labelling of the gel, \cite{debruynscaling}). 

The success of such a task is strongly conditioned by a better understanding of the physicochemical properties of the \trademark{Carbopol} gels which would allow developing versatile protocols of fluorescence labelling of the micro-gel particles which would allow the direct visualisation of the gel system under shear. 

From a theoretical perspective, our work clearly demonstrates the need of developing new frameworks to understand the unsteady yielding and slip in \trademark{Carbopol} gels. A modest progress in this direction has been already reported in Ref. \cite{solidfluid} where a simple model that can accurately describe the unsteady yielding in a rheometric setup in the absence of wall slip was proposed and validated agains experimental data. 

In spite its ability to describe a large amount of rheological data in both shear and oscillation and within a wide range of polymer concentrations and at various operating temperatures(see Refs. \cite{solidfluid, miguelstab, thermo}), this model is minimalistic and can be considered for now as \textit{"poor man's model"}: it is a scalar model, it has not been derived from first principles, it lacks a thermodynamical validation and, in its present form, it can not account for wall slip. 

Future theoretical work is needed to correct these deficiencies and to ultimately attempt to model realistic viscoplastic flows of industrial relevance such as flows around solid objects moving through a gel \cite{sedimentation}, unsteady start-up pipe flows and thermo convective flows.   

\section*{Acknowledgments}
The work was supported by the grant \textit{"ANR ThIM"} generously provided by the \textit{"Agence Nationale de la Recherche"}, France.
 
M. M. - G.  gratefully acknowledges the financial support provided by the \textit{\'{E}cole Polytechnique de l'Universit\'{e} de Nantes} and the hospitality of the \textit{Laboratoire de Thermocin\'{e}tique}, Nantes, France where a substantial part of the work has been completed. 
 
 We are grateful to Dr. Albert Magnin, Dr. Emmanuel Plaut and Dr. Cristel Metivier for many inspiring discussion on the rheology of \trademark{Carbopol} gels within the framework of the collaborative project \textit{"ANR ThIM"}.  
 
 T.B. acknowledges numerous insightful discussions with Professor Ian A. Frigaard on the yielding of \trademark{Carbopol} gels.

 A. P., T. B. and C. C. thank Christophe Le Bozec and J\'{e}r\^{o}me Delmas for their help in setting up the experiments.

We gratefully acknowledge the technical support provided by Mr. E. Roussel and Dr. P. Sierro from Thermo Fisher Scientific, Karlsruhe, Germany for the calibration of the nano-torque module installed on the Mars III rheometer and valuable discussions.

\bibliographystyle{elsarticle}

\providecommand{\noopsort}[1]{}\providecommand{\singleletter}[1]{#1}%

\end{document}